\newcommand{\ba}{BaCu$_2$Si$_2$O$_7$}

\documentstyle[prb,aps]{revtex}
\begin{document}
\input{psfig.sty}
\draft

\twocolumn[\hsize\textwidth\columnwidth\hsize\csname
@twocolumnfalse\endcsname

\title
{Spin dynamics in the $S=1/2$ weakly-coupled chains
antiferromagnet \ba.}

\author{A. Zheludev}
\address{Physics Department, Brookhaven National Laboratory, Upton, NY
11973-5000, USA.}

\author{M. Kenzelmann}
\address{Oxford Physics, Clarendon Laboratory, Oxford
OX1 3PU, UK.}

\author{S. Raymond}
\address{DRFMC/SPSMS/MDN, CENG, 17 rue des Martyrs, 38054
Grenoble Cedex, France.}

\author{T. Masuda and  K.
Uchinokura}
\address{Department of Advanced Material Science,  The
University of Tokyo, 7-3-1 Bunkyo-ku, Tokyo 113-8656, Japan.}

\author{S.-H. Lee}
\address{NIST Center for Neutron Research, National
Institute of Standards and Technology, MD 20899, USA.}

\date{\today}
\maketitle
\begin{abstract}
Inelastic neutron scattering is used to identify single-particle
and continuum spin excitations in the quasi-one-dimensional
$S=1/2$ antiferromagnet \ba. In the data analysis, close attention
is given to resolution effects. A gap in the continuum spectrum at
$\Delta_c=4.8$~meV is directly observed. The gap energy is in
excellent agreement with existing theoretical predictions. Below
the gap, the spectrum is dominated by sharp accoustic spin wave
excitations.
\end{abstract}


]

\narrowtext

\section{introduction}
Despite the apparent simplicity of the quantum $S=1/2$
antiferromagnetic (AF) Heisenberg model, the quest for a complete
understanding  of this system remains one of central problems in
magnetism. Semi-classical spin wave theory (SWT) is known to
provide an adequate description of the 3-dimensional (3D) version
of the model. After extensive theoretical and experimental studies
it is fair to say that  the purely one-dimensional (1D) case is
rather well understood as well. Its static and dynamic properties
totally defy the conventional spin wave picture. In 1D the ground
state looks nothing like the classical two-sublattice N\'{e}el
state, and is described as a ``marginal spin liquid'', with
power-law spatial spin correlations and no long-range
order.\cite{Bethe31} The spectrum contains no single-particle
magnon excitation, but  is instead a diffuse continuum described
in terms of two-particle ``spinon''
states.\cite{Fadeev81,Haldane93,Muller,Karbach97} Of great current
interest is the {\it quasi}-1 dimensional case of weakly
interacting quantum spin chains. Studies of such systems provide
the missing link between the exotic quantum-mechanical behavior of
the 1D model and the more familiar semi-classical spin wave
dynamics realized in three dimensions.

A great deal of experimental information on the quasi-1D
Heisenberg model was gained in studies of
KCuF$_3$.\cite{Satija80,KCUF3,Tennant95,Lake00} Particularly
exciting was the recent observation of a novel longitudinal
excitation, polarized along the direction of ordered moment, and
thus totally absent in SWT.\cite{Lake00} The recently
characterized \ba\
(Refs.~\onlinecite{Tsukada99,Zheludev00,Kenzelmann01}) is another
model weakly coupled chains compound, shown to be particularly
useful for studying transverse spin fluctuations, i.e., those
polarized perpendicular to the ordered moment. In a recent short
paper\cite{Zheludev00} we have demonstrated that the transverse
excitation spectrum in \ba\ has a unique {\it dual} nature. At
high energies the imaginary part of dynamic susceptibility is
dominated by continuum excitations, characteristic of the quantum
1D model. However, at energies below the characteristic strength
of inter-chain interactions, the continuum is ``cleaned up'' and
replaced by sharp single-particle states, similar to conventional
spin waves. A further analysis suggested that continuum
excitations appear only above a well-defined threshold energy.
Unfortunately, for purely technical reasons, this gap in the
continuum could not be observed directly. In the present paper we
report a detailed study of continuum excitations in the vicinity
of the energy gap.

\section{An effective spin Hamiltonian for \ba} \label{ham} Before we proceed to
describe the experimental procedures and results we shall briefly
review the structure and relevant magnetic interactions in \ba,
that are, by now, rather well documented. The silicate
crystallizes in an orthorhombic structure, space group
\textit{Pnma}, with lattice constants are $a = 8.862(2)\,$\AA, $b
= 13.178(1)\,$\AA \, and $c = 6.897(1)\,$\AA.\cite{Oliveira}
Magnetic properties are due to Cu$^{2+}$ ions that form
weakly-coupled antiferromagnetic chains running along the
crystallographic $c$-axis. Correspondingly, the temperature
dependence of bulk magnetic susceptibility follows the theoretical
Bonner-Fisher curve \cite{Bonner64} with an in-chain coupling
$J=24$~meV.\cite{Tsukada99} A similar estimate was obtained in
direct inelastic neutron scattering measurements of the 1D
zone-boundary energy $\hbar\omega_{\rm ZB}$ of magnetic
excitations, related to $J$ through $J=2\hbar\omega_{\rm
ZB}/\pi$.\cite{DCP} This measurement gives
$J=24.1$~meV.\cite{Kenzelmann01} Nearest-neighbor spin-spin
separation within the chains is equal to $c/2$, so the 1D AF zone
center $q_{\|}=\pi$ corresponds to $l=1$, where $\bbox{Q}=(h,k,l)$
denotes a wave vector in reciprocal space.

Weak interactions between the chains result in long-range
antiferromagnetic ordering at $T_{\rm N
}=9.2\;\mathrm{K}$.\cite{Tsukada99} The ordered moment is parallel
to the chain axis and extrapolates to a saturation value of
$0.15~\mu_{\rm B}$ at $T\rightarrow
0$.\cite{Tsukada99,Zheludev00,Kenzelmann01} In the ordered state
the alignment of nearest-neighbor spins is parallel along the $a$,
and antiparallel along the $b$ and $c$ directions, respectively.
The 3D AF zone-center is located at $\bbox{Q}=(0,1,1)$. The
relevant inter-chain exchange constants were determined from
measurements of spin wave dispersion along several directions in
reciprocal space.\cite{Kenzelmann01} The effective spin
Hamiltonian was established:
\begin{eqnarray}
 H  & = &\sum_{i,j,n}
 J\bbox{S}_{i,j,n}\bbox{S}_{i,j,n+1}\nonumber\\
 &+&  J_x\bbox{S}_{i,j,n}\bbox{S}_{i+1,j,n}
 +  J_y\bbox{S}_{i,j,n}\bbox{S}_{i,j+1,n}\nonumber\\
 &+&  J_3\bbox{S}_{i,j,n}\bbox{S}_{i+1,j+1,n}
 +  J_3\bbox{S}_{i,j,n}\bbox{S}_{i+1,j-1,n}\label{hamiltonian}.
\end{eqnarray}
Here the sum is taken over all spins in the system. The indexes
$i$ and $j$ enumerate the spin chains along the $a$ and $b$ axes,
respectively, and $n$ labels the spins within each chain. The
experimental values of inter-chain exchange parameter
are:\cite{Kenzelmann01}
\begin{eqnarray}
 J_x=-0.460(7)~\mathrm{meV},\nonumber\\
 J_y=0.200(6)~\mathrm{meV},\nonumber\\
 2J_3=0.152(7)~\mathrm{meV}\label{param}.
\end{eqnarray}

The spin chains in \ba\ are not straight, but slightly zig-zag.
The magnetic sites are displaced from high-symmetry
$(\case{1}{4},0,\case{1}{4})$ positions as described in
Ref.~\onlinecite{Tsukada99}. For the neutron scattering
measurements described below, this fact has one unfortunate
consequence. The 3D magnetic dynamic structure factor
$S(\bbox{Q},\omega)$ probed experimentally is not that of an ideal
straight chain, $S_0(\bbox{Q},\omega)$, but related to it through
the following rather complicated relation:
\begin{eqnarray}
&S&(\bbox{Q},\omega) =
 \cos^2(2\pi h\delta_x)\cos^2(2\pi k\delta_y)\cos^2(2\pi l\delta_z)S_0(\bbox{Q},\omega)\nonumber\\
 & + & \cos^2(2\pi h\delta_x)\cos^2(2\pi k\delta_y)\sin^2(2\pi l\delta_z)S_0(\bbox{Q}+(100),\omega)\nonumber\\
 & + & \cos^2(2\pi h\delta_x)\sin^2(2\pi k\delta_y)\cos^2(2\pi l\delta_z)S_0(\bbox{Q}+(011),\omega)\nonumber\\
 & + & \cos^2(2\pi h\delta_x)\sin^2(2\pi k\delta_y)\sin^2(2\pi l\delta_z)S_0(\bbox{Q}+(111),\omega)\nonumber\\
 & + & \sin^2(2\pi h\delta_x)\cos^2(2\pi k\delta_y)\cos^2(2\pi l\delta_z)S_0(\bbox{Q}+(001),\omega)\nonumber\\
 & + & \sin^2(2\pi h\delta_x)\cos^2(2\pi k\delta_y)\sin^2(2\pi l\delta_z)S_0(\bbox{Q}+(101),\omega)\nonumber\\
 & + & \sin^2(2\pi h\delta_x)\sin^2(2\pi k\delta_y)\cos^2(2\pi l\delta_z)S_0(\bbox{Q}+(010),\omega)\nonumber\\
 & + &
 \sin^2(2\pi h\delta_x)\sin^2(2\pi k\delta_y)\sin^2(2\pi l\delta_z)S_0(\bbox{Q}+(110),\omega)\label{3d}
\end{eqnarray}
In this equation $\delta_x\approx 0.028$, $\delta_y\approx0.004$
and $\delta_z\approx0.044$ are Cu$^{2+}$ displacements from
high-symmetry positions as described in
Ref.~\onlinecite{Tsukada99}.\footnote{We found that Eq. 8 in
Ref.~\protect\onlinecite{Tsukada99} is incorrect. Equation
\protect\ref{3d} is the corrected version of this formula,
rewritten to take $\delta_y$ into accounted.} It is important to
emphasize that Eq.~\ref{3d} is an exact result. At small wave
vectors only the first term is relevant and the dynamic structure
factor is that of a straight chain. However, at higher wave vector
transfers the contributions of other terms become significant.

\section{Experimental procedures}
The main goal of the present work was to reliably resolve
low-energy spin wave and continuum excitations near the 1D AF
zone-center in \ba.  Previous studies indicated that in \ba\ the
continuum sets in between 3.5~meV and 5.5~meV energy
transfer.\cite{Zheludev00} The most interesting dynamic range is
therefore between zero and 10~meV. The choice of spectrometer
configurations in inelastic neutron scattering experiments was
therefore governed by the conflicting needs of intensity, high
wave vector resolution along the chain axis (required to overcome
the steep dispersion along the chains), energy resolution, and the
need to cover a wide range of energy transfers. Since it appears
impossible to simultaneously satisfy all these conditions, several
different experimental setups with complementary characteristics
were utilized. Quantitative comparisons between data sets
collected in different configurations was essential to verifying
the validity of the results. In all cases we used the same \ba\
single-crystal sample as in previous
studies.\cite{Zheludev00,Kenzelmann01} In all configurations the
crystal was mounted with the $(0,k,l)$ reciprocal-space plane in
the scattering plane of the spectrometer.

Ideally, the measurements are done around $\bbox{Q}=(0,0,1)$
($q_\|=\pi$), where the extra terms in Eq.~\ref{3d} are
negligible, and all magnetic scattering is due to spin
fluctuations transverse to the direction of ordered moment ($c$
axis). Unfortunately, at this wave vector, certain geometric
constrains restrict the accessible range of energy transfers to
just over 4~meV. Measurements in this range were performed on the
IN14 cold-neutron spectrometer installed at Institute
Laue-Langevin (ILL), using a very high energy and wave vector
resolution configuration (Setup I): ${\rm guide}-40'-40'-{\rm
open}$ collimations, $E_{\rm f}=3$~meV fixed final-energy
neutrons, and a Be filter after the sample. To cover a wider
energy-range and still have reasonable wave vector and energy
resolutions we employed IN14 in ``anti-W'' geometry with ${\rm
guide}-40'-60'-{\rm open}$ collimations and final neutron energy
fixed at $E_{\rm f}=5$~meV (Setup II). The data were collected
around $\bbox{Q}=(0,0,3)$ ($q_\|=3\pi$), where the contributions
of both the first and second terms in Eq.~\ref{3d} are
significant. In this geometry measurements could be extended up to
8~meV energy transfer. A similar configuration was used in
combination with a horizontally-focusing analyzer (Setup III).
While in this latter mode the wave vector resolution is
considerably worse than with a flat analyzer, the intensity gain
is quite substantial. Energy transfers of up to 17~meV were
achieved using the IN22 spectrometer at ILL, with ${\rm
guide}-60'-60'-{\rm open}$ collimations and final neutron energy
fixed at $E_{\rm f}=14.7$~meV (Setup IV). In this case a Pyrolitic
Graphite (PG) filter was used after the sample.

\begin{figure}
\psfig{file=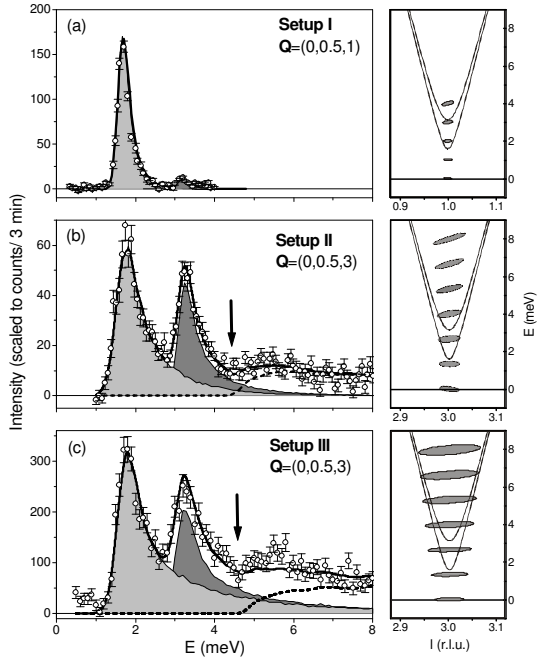,width=3.2in,angle=0}\caption{
 Left: Typical constant-$Q$ scans collected in \ba\ at the 1D AF
 zone-center using different spectrometer configurations. Heavy solid
 lines represent a semi-global fit to the data as described in the
 text. Shaded areas are contributions of single-particle
 excitations given by Eqs.~\protect\ref{Sx},\protect\ref{Sy}. Dashed lines show the continuum
 contribution, as expressed by Eq.~\protect\ref{cont}. Arrows indicate the slight dip in the observed intensity that
 corresponds to the continuum energy gap $\Delta_c$. Right:
 Evolution of the calculated FWHM resolution ellipsoids in the
 course of the corresponding scans, plotted in projection onto the $(l,\hbar \omega)$ plane.
 Solid lines represent the spin wave dispersion relation.}
 \label{qdata1}
\end{figure}

In all experiments sample environment was a standard ``ILL
Orange'' He-4 cryostat and all data were collected at either
$T=1.5$~K or $T=2$~K. For the cold neutron experiments the
background was measured in constant-$Q$ scans, away from the 1D AF
zone-center, where no magnetic scattering is expected to occur,
due to a step dispersion along the chain axis. In all cases the
background was found to be completely flat and featureless,
typically 2.5, 4 and 5 counts per minute for setups I, II and III,
respectively. The measured background was fit to a straight line
and is subtracted from all the data sets shown below. For the
thermal neutron experiment (Setup IV) the background was measured
on an empty sample can and subtracted from the experimental data
in a similar fashion.

\begin{figure}
\psfig{file=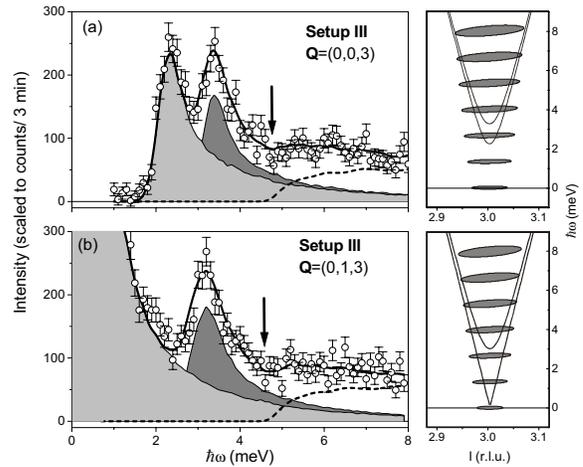,width=3.2in,angle=0}\caption{
 Constant-$Q$ scans collected in \ba\ at the 1D AF
 zone-center at different momentum transfers perpendicular to the chains using Setup III.
 Lines, shaded areas and graphical representations of the resolution functions on the left are as in
 Fig.~\protect\ref{qdata1}.}
\label{qdata2}
\end{figure}

\section{Results}
\subsection{Transverse spin excitations}
%For momentum transfers almost parallel to the direction of ordered
%moments ($c$ axis) the contribution of longitudinal excitations is
%negligible and one directly measures the transverse-polarized
%spectrum.
%
Representative constant-$Q$ scans measured at the 1D AF
zone-center $q_\|=\pi,~3\pi$ using Setups I, II and III are shown
in Fig.~\ref{qdata1}. To the right of each scan we plot the
calculated FWHM resolution ellipsoids in projection on the
$(l,\hbar\omega)$ plane to show their size and orientation
relative to the spin wave dispersion branches originating from the
1st and 2nd terms in Eq.~\ref{3d} (solid lines). Three distinct
components of the excitation spectrum are apparent. The sharp peak
at $\hbar \omega\approx 2$~meV is the single-particle spin wave
mode given by the first term in Eq.~\ref{3d}. This feature
dominates the low-energy spectrum around $\bbox{Q}=(0,0,1)$, where
the other terms in Eq.~\ref{3d} are negligible thanks to the small
value of corresponding trigonometric prefactors (Fig.~\ref{qdata1}
(a) and Figs.~1,2 in Ref.~\onlinecite{Zheludev00}). At $q_\|=3\pi$
however, the second term in Eq.~\ref{3d} becomes significant and
gives rise to the 3~meV peak in Figs.~\ref{qdata1} (b) and (c). In
essence, this second peak is an ``echo'' of the same spin wave
excitation visible by virtue of a non-trivial 3D arrangement of
magnetic sites. The third feature in Fig.~\ref{qdata1} is the
diffuse scattering above the spin wave peaks that, as will be
discussed in detail below, is not a resolution-enhanced ``tail''
of the spin wave peaks, but a separate entity. This scattering is
attributed to continuum excitations in the system. In
Figs.~\ref{qdata1} (b) and (c) one can see a weak yet reproducible
intensity dip that separates this continuum scattering from the
underlying spin waves (arrows).

\begin{figure}
\psfig{file=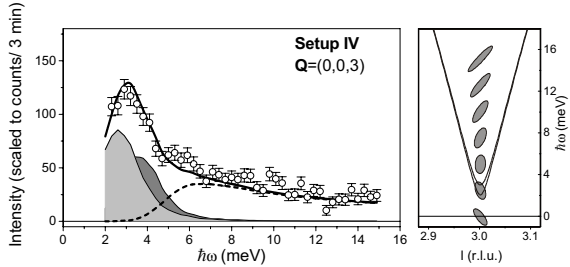,width=3.2in,angle=0}\caption{
 Constant-$Q$ scan collected in \ba\ at the 1D AF
 zone-center using the thermal-neutron Setup IV.
 Lines, shaded areas and graphical representation of the resolution function  on the left are as in
 Fig.~\protect\ref{qdata1}.}
\label{qdata3}
\end{figure}

Figure.~\ref{qdata2} shows constant-$Q$ scans collected with Setup
III at different wave vector transfers perpendicular to the
chains. The dispersion of the lower-energy spin wave is quite
apparent. An almost total absence of dispersion in the 3~meV peak
is consistent with that it originates from $\bbox{Q}+(100)$
scattering (Eq.~\ref{3d}), and the known spin wave dispersion
along the $(1,k,1)$ reciprocal-space rod
(Ref.~\onlinecite{Kenzelmann01}, Fig.~6). In all cases the
continuum is observed, and the intensity dip occurs at roughly the
same energy (arrows). There seems to be no obvious wave vector
dependence of the continuum intensity along the $b^{\ast}$
direction. The energy resolution of the thermal-neutron experiment
(setup IV) is too coarse to resolve the two spin wave branches and
the continuum. Figure.~\ref{qdata3} shows an energy scans
collected in this configuration at $\bbox{Q}=(0,0,3)$.

\begin{figure}
\psfig{file=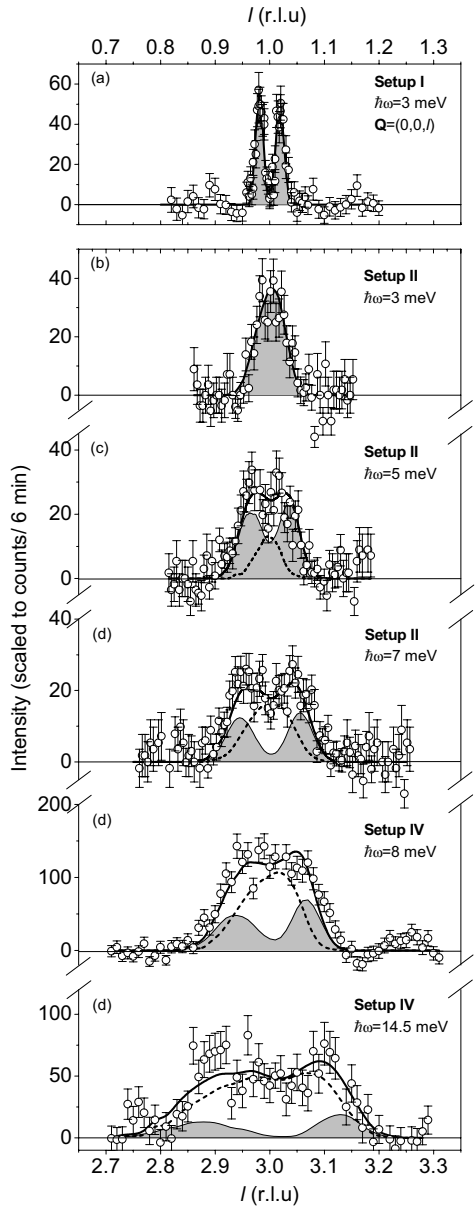,width=3.2in,angle=0}\caption{
 A series of constant-$E$ scans along the spin chains in \ba.
 Lines and shaded areas are as in
 Fig.~\protect\ref{qdata1}.}
\label{edata}
\end{figure}

Several constant-$E$ scans measured with different configurations
are shown in Fig.~\ref{edata}. At 3~meV energy transfer
(Fig.~\ref{edata}(a) ) two spin wave peaks are perfectly resolved
using the highest-resolution configuration (Setup I). This proves
that at these energies the excitation spectrum is entirely due to
single-particle excitations. At the same energy transfer Setup II
lacks the wave vector resolution to resolve the spin wave peaks,
as can be seen from Fig.~\ref{edata} (b). However, as will be
discussed in detail in the next section, at higher energies
(Fig.~\ref{edata} (c,d)) while the resolution is sufficient to
resolve individual spin wave peaks, only a single broad feature is
observed. As explained in Ref.~\onlinecite{Zheludev00}, this
behavior is a result of the excitation continuum ``filling in''
the space between the spin wave branches. The trend continues at
even higher energies, as can be seen from Fig.~\ref{edata} (e) and
(f), that shows scans collected using Setup IV.

\subsection{Data analysis}
Resolution effects, particularly wave vector resolution along the
chain axis, are clearly a major factor in the experiments
discussed here. In order to reliably distinguish between continuum
scattering and the high-energy resolution ``tails'' of the
single-particle peaks, one has to de-convolute the signal and the
resolution function of the spectrometer. In practice this is done
by constructing a parameterized model cross section, numerically
convoluting it with the calculated resolution, and adjusting the
parameters to best-fit the data. Confidence in this procedure can
be gained by {\it simultaneously} fitting data sets collected in
spectrometer configurations with different resolution
characteristics, while using the {\it same} model cross section
function and parameters.

\subsubsection{Model cross section function}
A simple yet accurate description of spin dynamics in
weakly-interacting Heisenberg chains is provided by the Mean
Field/ Random Phase Approximation (MF/RPA) treatment of
inter-chain interactions.\cite{Schulz96,Essler97} Within this
framework, the $T=0$ transverse-polarized spectrum of a
weakly-ordered quasi-1D $S=1/2$ system has two distinct
contributions.

\begin{figure}
\psfig{file=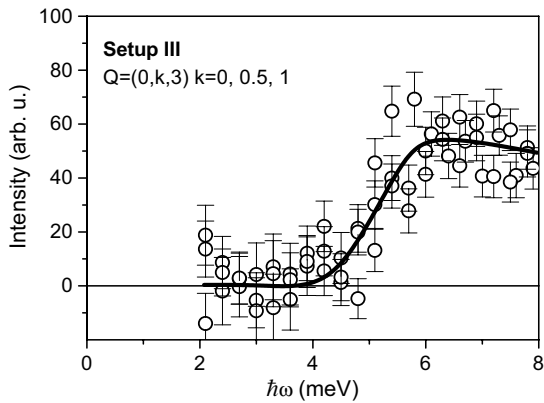,width=3.2in,angle=0}\caption{
 Measured intensity of transverse-polarized continuum scattering in \ba, obtained by subtracting the
 simulated single-mode contribution from constant-$Q$ scans as explained in the text. The solid
 line is a guide for the eye.}
\label{purecont}
\end{figure}

The first contribution is from transverse-polarized
single-particle excitations (spin waves), associated with the
presence of long-range antiferromagnetic order. For \ba\ the spin
wave dispersion has been investigated in some
detail\cite{Kenzelmann01} and found to be in excellent agreement
with MF/RPA theoretical predictions. In the present work we shall
employ the same expression for the spin wave structure factor, as
in Ref.~\onlinecite{Kenzelmann01}:
 \begin{eqnarray}
 S^{xx}_{{\rm SM},0}(\bbox{Q},\omega) &= &
  A\frac{[1- \cos(\pi l)]}{2}\delta\left[\omega^2-\omega_x^2(\bbox{Q})\right],\label{Sx}\\
 S^{yy}_{{\rm SM},0}(\bbox{Q},\omega)& = &
  A\frac{[1- \cos(\pi l)]}{2}\delta\left[\omega^2-\omega_y^2(\bbox{Q})\right],\label{Sy}\\
 \left[ \omega_{x,y}(\bbox{Q})\right]^2 & =& \frac{\pi^2}{4}J^2\sin^2(\pi l) \nonumber\\
 & + & \frac{\Delta^2}{|J'|}  \left[|J'|+ J(\bbox{Q})\right]+ D_{x,y}^2,\\
\label{disp}\label{dd}
 |J'|& \equiv & \frac{1}{2}( |J_{\rm x}|+|J_{\rm y}|+2|J_{\rm
3}|).
\end{eqnarray}
In this formula $S^{xx}_{{\rm SM},0}(\bbox{Q},\omega)$ and
$S^{xx}_{{\rm SM},0}(\bbox{Q},\omega)$ are single-mode (SM)
dynamic structure factors for excitations polarized along the $a$
and $b$ axes, respectively, assuming {\it straight} spin chains in
the system. The actual SM structure factors $S^{xx}_{{\rm
SM}}(\bbox{Q},\omega)$ and $S^{yy}_{{\rm SM}}(\bbox{Q},\omega)$
for the zig-zag chain in \ba\ are related to $S^{xx}_{{\rm
SM},0}(\bbox{Q},\omega)$ and $S^{yy}_{{\rm
SM},0}(\bbox{Q},\omega)$ through Eq.~\ref{3d}. The parameter $A$
in Eq.~\ref{disp} represents an overall intensity prefactor, while
$D_x=0.36$~meV and $D_y=0.21$~meV\cite{Kenzelmann01} are anisoropy
gaps for the two spin wave branches. In Eq.~\ref{disp}
$J(\bbox{Q})$ is the Fourier transform of inter-chain coupling,
and is given by
\begin{eqnarray}
&J(\bbox{Q})=2J_{x}\cos(\pi h)+2J_{y}\cos(\pi k)+ & \nonumber
\\ & 2J_{3}\cos(\pi(h+k))+2J_{3}\cos(\pi(h-k))&.\label{J}
\end{eqnarray}

The ``mass gap'' $\Delta$ in Eq.~\ref{disp} defines the bandwidth
of spin wave dispersion perpendicular to the chains. Its physical
meaning deserves some comment. In the chain-MF model, in the
magnetically ordered state, each spin chain experiences an
effective staggered field $H_{\pi}$ generated by the static
staggered magnetization in adjacent chains.  The staggered field
produces a confining potential between spinons in an $S=1/2$
chain, yielding a two-spinon bound state.\cite{Schulz96} This new
single-particle excitation has a gap $\Delta$ that scales as
$\Delta \propto H_{\pi}^{2/3}$.\cite{Oshikawa97} When inter-chain
interactions are switched on at the RPA level, the bound state
acquires a dispersion perpendicular to the chain axis and takes
the role of a conventional spin wave. At the 3D AF zone-center
(the location of magnetic Bragg reflections) its energy goes to
zero, as appropriate for a Goldstone mode. The mass gap $\Delta$
still enters the dispersion relation, as in Eq.~\ref{disp}. It is
directly related to the strength of inter-chain interactions
through:\cite{Schulz96}
\begin{equation}
 \Delta  \approx 6.175 |J'|\label{Delta}.
\end{equation}
From the known values of inter-chain coupling constants for \ba\
one gets\cite{Kenzelmann01} $\Delta=2.51$~meV. \footnote{It has
recently been suggested that the numerical coefficient in
Eq.~\protect\ref{Delta} may be incorrect.\protect\cite{Essler} The
actual value of this coefficient is actually irrelevant for this
work. Indeed, the microscopic exchange constants $J_1$--$J_3$ were
deduced from the measured spin wave  bandwidths in \ba\ using the
{\it same} value of the coefficient. Note that it is $\Delta$, and
not the actual $J$-values, that determine the spin wave dispersion
perpendicular to the chains in Eq.~\protect\ref{dd}. Unlike the
exchange parameters, that characterize the {\it model
Hamiltonian}, $\Delta$ is a true observable physical quantity,
equal to the measurable spin wave energy at
$\bbox{Q}=(0.5,0.5,1)$, where inter-chain interactions in \ba\
cancel out at the RPA level, and is thus independent of any
theoretically-derived constants.  }

The second contribution to the transverse dynamic structure factor
in the MF/RPA model is what remains of the 2-spinon continuum in
isolated chains.\cite{Essler97,Schulz96} Unfortunately, a
convenient analytical expression for this diffuse part of cross
section, one that could be used to analyze the data, is not
currently available. In our analysis we instead employed a simple
ansatz that has the same main characteristics as the MF/RPA
result: i) At high energies the Muller-ansatz form that describes
2-spinon continuum scattering in isolated
chains\cite{Mulleransatz} is recovered; ii) The continuum has a
sharp step-function low-energy cutoff (gap) $\Delta_c=2\Delta$ at
the 1D AF zone-center $q_\|=\pi$; iii) Unlike the spin waves, the
lower continuum bound shows no dispersion perpendicular to the
chain axis; iv) The $q_\|$- dependence of the lower bound is given
by $[\Delta_c(q_\|)]^2=\Delta_c^2+\case{\pi^2}{4}J^2\sin^2(q_\|)$;
and v) Except at ``magic'' reciprocal-space point $(0.5,0.5,1)$,
where inter-chain interactions cancel out at the RPA level, the
continuum has no singularity at the lower bound. To analyze our
data, where none of the measured scans were taken in the vicinity
of the magic wave vector, we used the following empirical function
to model the continuum part of the dynamic structure factor:
\begin{eqnarray}
  S^{xx,yy}_{{\rm c},0}(\bbox{Q},\omega)
 &  = &\alpha A \frac{[1- \cos(\pi l)]}{2}\nonumber\\
 & \times & \frac{1}{
 \sqrt{\omega^2-\case{\pi^2}{4}J^2\sin^2(q_\|)}}
 ~\theta\left[\omega-\omega_c(\bbox{Q})\right]\label{cont}\\
 \left[ \omega_{c}(\bbox{Q}) \right]^2& =& \Delta_c^2+\frac{\pi^2}{4}J^2\sin^2(\pi
  l)\label{contdisp}
\end{eqnarray}
The parameter $\alpha$ is the intensity of continuum scattering
relative to that of the spin waves,  and $\theta(x)$ is the
Heavyside step function. In essence, Eq.~\ref{cont} is simply the
M\"{u}ller ansatz truncated at the expected lower continuum bound.
For the 3D spin arrangement in \ba\ the actual continuum cross
sections $S^{xx,yy}_{{\rm c}}$ are related to $S^{xx,yy}_{{\rm
c},0}$ through Eq.~\ref{3d}.

\subsubsection{Choice of adjustable parameters}
The goal was to perform a global analysis of all data sets
collected in different configurations using as few adjustable
parameters as possible. All relevant exchange constants and
anisotropy parameters have been measured
previously\cite{Kenzelmann01}. These constants were fixed to the
values quoted in Sec.~\ref{ham}. Since we have not performed
absolute callibration of intensities measured in various setups, a
separate value of the intensity prefactor $A$ was used for each
configuration. However, the same values of $A$  were used for
different scans measured using the same setup. In our analysis the
continuum gap $\Delta_c$ was used as a separate variable,
independent of $\Delta$. As with the continuum intensity prefactor
$\alpha$, the same value of $\Delta_c$ was used for all data sets
measured in all configurations.

In Eq.~\ref{3d} only the first two terms were found to be of any
significance for the particular experiment, where the data were
collected at small momentum transfers along the  $a$- and $b$
axes. Equation~\ref{3d} could thus be greatly simplified by
setting the transverse zig-zag displacements $\delta_x$ and
$\delta_y$ to zero. The relevant longitudinal displacement
parameter $\delta_z$ can be expected to have a  slight temperature
dependence. Only the room-temperature value $\delta_z=0.044$ is
known from structural data,\cite{Oliveira} and in our analysis it
had to be treated as an adjustable variable. The same value of
$\delta_z$ was used for all data sets simultaneously.

\subsubsection{Fits to experimental data}
All together, 9 constant-$Q$ and 8 constant-$E$ scans were
analyzed, including all those shown in
Figs.~\ref{qdata1}--\ref{edata}.  The model cross section, defined
by Eqs.~\ref{disp}--\ref{contdisp} and Eq.~\ref{3d} was
numerically folded with the calcualted Cooper-Nathans resolution
functions\cite{Cooper67} for each setup. The calculation was based
on the known geometry of the spectrometers and characteristics of
the neutron guides, monochromator and analyzer crystals, and
collimators used in the experiments.  The reliability of the
resolution calculation was checked against the measured energy
widths of elastic incoherent scattering from the sample in each
setup. The folded cross section was scaled by the square of the
magnetic form factor of Cu$^{2+}$ and  fit to the data using a
least-squares algorithm.

The result of the fit ($\chi^2=1.7$) is shown in solid lines  in
Figs.~\ref{qdata1}--\ref{edata}. The shaded areas are
contributions of the single-mode part of the cross section. In the
constant-energy scans the contributions of the spin wave branches,
originating from the first two terms in Eq.~3d, are merged into a
single profile. These two contributions are shown in shading of
different intensity in the constant-$Q$ data. The dashed lines
represent the continuum part of the cross section. The following
fit parameters were obtained: $\Delta_c=4.8(2)$~meV,
$\alpha=0.41(3)$ and $\delta_z=0.041(1)$.

It is gratifying to see that the model fit function reproduces the
available data extremely well with just a few parameters. The
refined value for $\delta_z$ is very close to its room-temperature
value. The fitted continuum gap $\Delta_c$ is within the error bar
of the chain-RPA result: $\Delta_c=2\Delta=5.0$~meV.

To extract information on the excitation continuum without having
to rely on a specific ansatz to describe its shape, we employed an
alternative data analysis procedure. The data collected in
constant-$Q$ scans using Setup III below 3.5~meV were fit using
only the SM part of the model cross section function
($\chi^2=1.2$). The result was then used to simulate the SMA
contribution in the entire scan range and subtracted from the
measured data. The remaining intensity is due to continuum
scattering and is shown in Fig.~\ref{purecont}, where the results
obtained at $\bbox{Q}=(0,0,3)$, $(0,0.5,3)$ and $(0,1,3)$ have
been averaged. In this plot the threshold at around $2\Delta$~meV
becomes clearly apparent.

\subsection{Attempt to observe the longitudinal mode.}
One of the most interesting predictions of the MF-RPA model for
weakly-coupled $S=1/2$ Heisenberg chains is the so-called
longitudinal mode: a single-particle excitation polarized along
the direction of ordered moment. This unique feature is totally
absent in conventional spin wave theory, where the excitations
represent the {\it precession} of the order parameter around its
equilibrium direction, and are thus transverse-polarized. The
longitudinal mode has been recently observed in the quasi-1D
$S=1/2$ system KCuF$_3$ by means of unpolarized\cite{Lake00} and
polarized\cite{Nagler} inelastic neutron scattering. In addition
to the study of transverse-polarized excitations in \ba, we
performed a dedicated experiment to search for the longitudinal
mode in this system.

\begin{figure}
\psfig{file=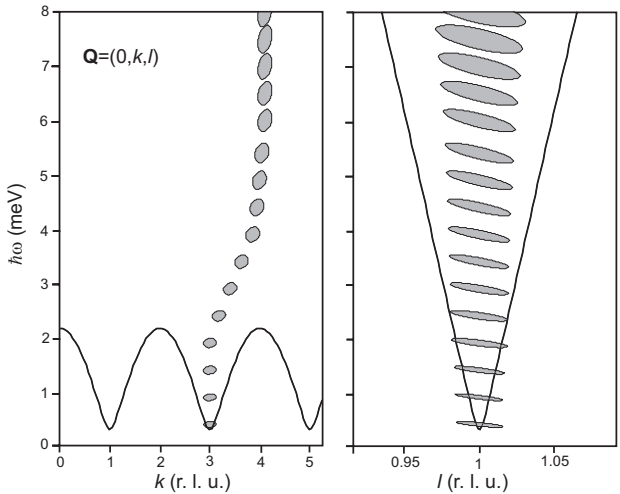,width=3.2in,angle=0}\caption{Reciprocal
space/ energy- trajectory of the inelastic scan measured in \ba\
using setup V, showing the orientation of FWHM resolution
ellipsoids relative to the spin wave dispersion.} \label{trajec}
\end{figure}
MF-RPA theory makes specific predictions for the energy and
intensity of the longitudinal mode. Unlike the transverse spin
waves that have a substantial dispersion perpendicular to the
chains and reach zero energy at the 3D AF zone-center, the
longitudinal mode is gapped and has very little dispersion in this
direction. Its RPA dynamic structure factor can be written
as:\cite{Essler97}
 \begin{eqnarray}
 S^{zz}_{{\rm SM},0}(\bbox{Q},\omega) &= &
  A\frac{\gamma}{2}\frac{[1- \cos(\pi l)]}{2}\delta\left[\omega^2-\omega_z^2(\bbox{Q})\right],\label{Sz}\\
 \left[ \omega_{z}(\bbox{Q})\right]^2 & =& \frac{\pi^2}{4}J^2\sin^2(\pi l) \nonumber\\
 & + & \frac{\Delta^2}{|J'|}  \left[3|J'|+ J(\bbox{Q})\right],\\ \label{long}
 \gamma & \approx\ 0.49
\end{eqnarray}
From this equation we see that the longitudinal mode is expected
to have a gap of $\sqrt{3}\Delta$ and be approximately four times
weaker in intensity than the transverse-polarized spin waves.

\begin{figure}
\psfig{file=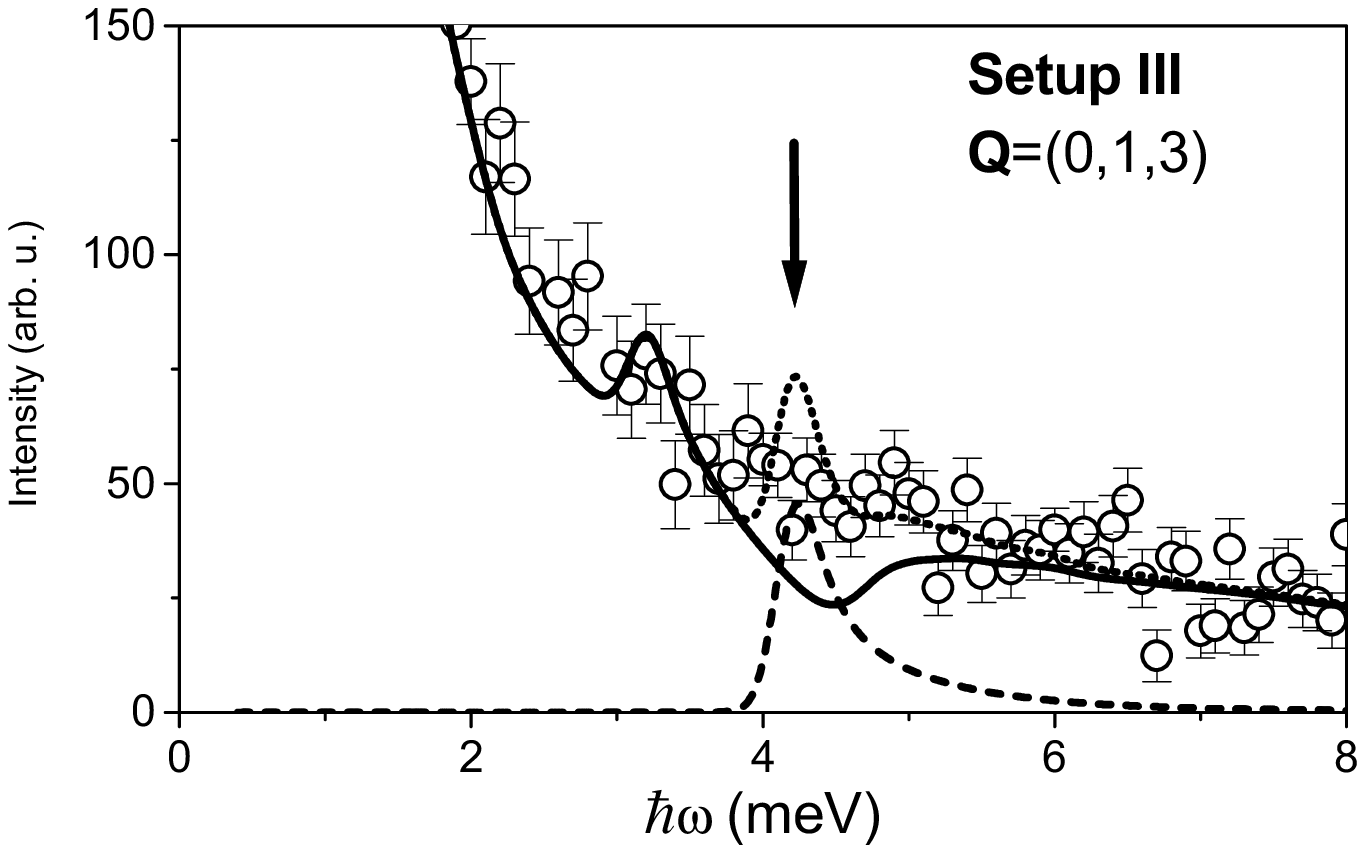,width=3.2in,angle=0}\caption{
 Constant-$q_\|$ scan collected in \ba\ at the 1D AF
 zone-center and large momentum transfer perpendicular to the chains, using Setup V.
 The trajectory of the scan is as shown in Fig.~\protect\ref{trajec}. Solid line
 is an estimate of the contribution of spin fluctuation perpendicular to the direction
 of ordered moment. The dashed line shows the expected contribution of the longitudinal
 mode, as predicted by the MF/RPA model.\protect\cite{Essler97} The dotted line is the
 MF/RPA prediction for the total scattering intensity.}
\label{spinsdata}
\end{figure}

To look for the longitudinal mode in \ba, measurements were
performed at large momentum transfers perpendicular to the chain
axis. In this geometry the angle between the ordered moment
($c$-axis) and the scattering vector is large, and both
longitudinal and transverse excitations contribute to scattering.
The data were collected at the NG5 cold 3-axis spectrometer at the
National Institute of Standards and Technology Center for Neutron
Research. The large momentum transfers perpendicular to the chains
enabled us to take full advantage of a horizontally-focusing PG
analyzer. The chain axis was at all times aligned as closely
parallel to the scattered beam as possible, ensuring that energy
and $c$-axis wave-vector resolution widths remained at least as
narrow as in the standard flat-analyzer mode. The data were thus
collected along a rather non-trivial trajectory in $E-Q$ space, as
illustrated in Fig.~\ref{trajec}. At all times the momentum
transfer along the chains was maintained at the 1D AF zone-center
$l=1$. Neutrons of $E_f=5$~meV were used with a Be filter
positioned after the sample and
$\mathrm{guide}-80'-80'(radial)-\mathrm{open}$ collimations (Setup
V). The background (flat, 3.4 counts/ min) was measured by
repeating the scan with the sample rotated by roughly 20$^{\circ}$
around the $a$ axis.

The data are shown in Fig.~\ref{spinsdata}. At low energies, below
3~meV energy transfer, we expect the main contribution to the
scattering intensity to originate from the gapless
transverse-polarized spin waves. After the inclusion of
appropriate neutron polarization factors, the cross section for
transverse excitations described above was fit to this low-energy
portion of the scan (up to 3~meV transfer). All parameters,
including $\Delta_c$ and $\alpha$ were fixed at the values
determined previously, and only the intensity prefactor was
treated as an adjustable variable. The result of the fit is shown
in a solid line in Fig.~\ref{spinsdata}. We see that the
transverse-polarized cross section alone reproduces the measured
scan rather well. The only inconsistency is the lack of a dip or
even inflection point in the measured intensity at $\Delta_c$.
However, the fit shows that experimental energy resolution should
have been sufficient to resolve it.

\begin{figure}
\psfig{file=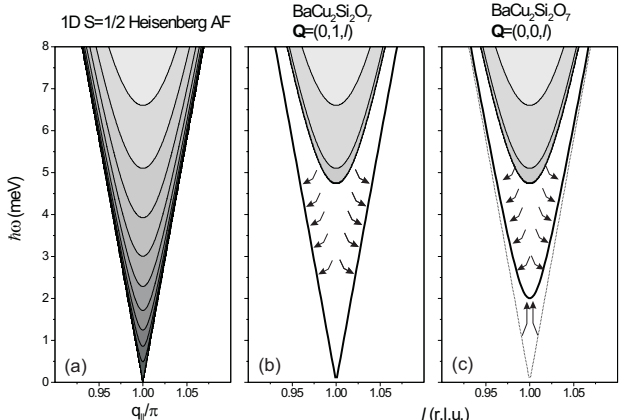,width=3.2in,angle=0}\caption{
 Re-distribution of spectral weight in a 1D $S=1/2$ Heisenberg AF (a) that occurs when inter-chain interactions are switched on, as in \ba\ (b and c). }
\label{cartoon}
\end{figure}

In principle, one could subtract the fit from the experimental
data and claim that the residual intensity is due to longitudinal
spin fluctuation. Given the severity of resolution effects,  we
were reluctant to push the data analysis to this level. What we
did instead was simulate the expected contribution from the
longitudinal mode, as given by Eq.~\ref{long} and an appropriate
neutron polarization factor, and shown in a dashed line in
Fig.~\ref{spinsdata}. Note that the ``extra'' intensity (arrow in
Fig.~\ref{spinsdata}), not accounted for by the transverse dynamic
structure factor, is centered at roughly the same energy as the
expected longitudinal excitation. Moreover, the intensity
``surplus'', integrated in the range 3.5--5.5~meV (3.0(0.4)
counts/(min meV)) is within the error bar of the expected
intensity of the longitudinal mode integrated in this range (2.7
counts/(min meV)). If the longitudinal mode was a sharp
single-particle excitation, as predicted by the RPA, a well
defined peak centered at 4.2~meV should have been easily observed
with our experimental statistics. This is illustrated by the
dotted line, which is the calculated sum of the transverse
contribution obtained in the fit and that given by Eq.~\ref{long}
for the longitudinal mode. The result can be reconciled with the
RPA model, if one assumes that the longitudinal mode has a finite
lifetime, i.e., an intrinsic energy width of roughly 1~meV. This
scenario would explain both the ``extra'' intensity and the
absence of a sharp peak. It is also consistent with experimental
studies of  KCuF$_3$,\cite{Lake00} where the observed longitudinal
mode {\it is} broadened and has an intrinsic width of roughly 1/4
of its central energy.

\section{Discussion and concluding remarks}
Despite the obvious difficulties associated with wave vector- and
energy resolution, the \ba\ data presented above contain
compelling evidence for a gap, or at least a {\it pseudo-gap}, in
the transverse-polarized continuum. The gap value
$\Delta_c=4.74.8$~meV$\approx2\Delta$ is in excellent agreement
with predictions of the MF-RPA model. Interestingly, non-linear
spin wave theory (NSWT) \cite{NSWT} predicts a much larger
threshold energy. In this model the lowest-energy
transverse-polarized continuum excitations are 3-magnon states,
and a pseudogap occurs at roughly $3\Delta$. An NSWT calculation
for \ba\ based on the known exchange parameters gives
$\Delta_c\approx 7.5$~meV.\cite{Zheludev00}

Figure~\ref{cartoon} is a cartoon illustration of how the spectral
weight of transverse-polarized excitations  gets re-distributed in
a 1D $S=1/2$ Heisenberg AF when inter-chain interactions are
switched on. The singularity on the lower bound in the 1D system
(Fig.~\ref{cartoon} (a)) becomes separated from the bulk of the
two-spinon continuum (Fig.~\ref{cartoon} (b) and (c)). The
spectral weight in its vicinity is consolidated (arrows) to yield
a sharp spin wave excitation (solid lines).

Our results regarding longitudinal excitations are less
conclusive. Clearly, additional measurements, possibly using
polarized neutrons, will be required to fully resolve this problem
for \ba. The available unpolarized data shown above are, in fact,
consistent with the MF-RPA model, but suggest a finite lifetime
for the longitudinal mode, as in KCuF$_3$.\cite{Lake00}

\acknowledgements We would like to thank I. Tsukada, Y. Sasago and
K. Kakurai for cooperation in earlier stages of this project, Dr.
L.~P. Regnault (CEA Grenoble) and Dr. A. Wildes for their
assistance with experiments at ILL, and Prof. F. H. L. Essler
(University of Warwick), Prof. R.~A. Cowley (Oxford University)
and Dr. I. Zaliznyak (BNL) for illuminating discussions. This work
is supported in part by the U.S.-Japan Cooperative Program on
Neutron Scattering, Grant-in-Aid for COE Research ``SCP coupled
system" of the Ministry of Education, Science, Sports, and
Culture. Work at Brookhaven National Laboratory was carried out
under Contract No. DE-AC02-98CH10886, Division of Material
Science, U.S.\ Department of Energy. One of the authors (M. K.)
was supported by a TMR-fellowship of the Swiss National Science
Foundation under contract No. 83EU-053223. Studies at NIST were
partially supported by the NSF under contract No. DMR-9413101.

%\bibliographystyle{prsty}
%\bibliography{nhaldane}

\end{document}